\documentclass[11pt,conference,comsoc]{IEEEtran}
\usepackage[T1]{fontenc}

\usepackage{graphicx,url}

\usepackage{cite}

\ifCLASSINFOpdf
\else
\fi
\usepackage{amsmath}
\usepackage[cmintegrals]{newtxmath}
\begin{document}
%
\title{ECN based Congestion Control for a Software Defined Network}
%
%
%

\author{\textsf{Mohana Prasad Sathya Moorthy,
        Manoj Kumar Sure,
        Krishna M. Sivalingam} \\
\textit{Dept. of CSE, Indian Institute of Technology Madras, Chennai,
  India, 600036} \\
}

\maketitle

\begin{abstract}

This paper deals with congestion control in a software defined network
(SDN) setting. Presently, explicit router schemes, such as Explicit
Congestion Notification (ECN), work in conjunction with the TCP
protocol to handle congestion in a distributed manner.  With the
emergence of SDN and centralized control, it is possible to leverage
the global view of the network state to make better congestion control
decisions. In this work, we explore the advantages of bringing in
global information into distributed congestion control. We propose a
framework where the controller with its global view of the network
actively participates in the congestion control decisions of the end
TCP hosts, by setting the ECN bits of IP packets appropriately. Our
framework can be deployed very easily without any change to the end
node TCPs or the SDN switches. We also show 30x improvement over the
TCP Cubic variant and 1.7x improvement over TCP/RED in terms of flow
completion times for one implementation of this framework, using the
Mininet emulator.
\end{abstract}


%
\IEEEpeerreviewmaketitle

\section{Introduction}

This paper deals with congestion control in computer networks.
Existing solutions can be categorized as end-node based or
router-based. The latter solutions use queue management and scheduling
algorithms that provide signals to the end hosts, to reduce the source
traffic. Active Queue Management techniques drop/mark packets at the
switch/router buffers thereby signalling the end nodes about
congestion \cite{feng,FloydRED,floyd3}. There are some schemes that use both
active queue management and end node TCP modifications
\cite{alizadeh,katabi}.


Most of the existing congestion control algorithms \cite{Winstein,
  Dong} have a very limited view of the network and its
traffic. Many TCP based congestion control algorithms use packet loss
as an indicator of congestion.  Another measure used is the round-trip
time (RTT). The problem with using RTT is that, the feedback may be
easily misinterpreted. For, example consider a 100 packet backlog
(with 1,500 Byte packets) in a router queue. It corresponds to
$1,200~\mu s$ of queuing delay at 1~Gbps, but only $120~\mu s$ at
10~Gbps. The end node cannot make fine distinctions without more
information. Without the detailed knowledge of the underlying network,
TCP will continuously keep increasing and decreasing its congestion
window trying to adapt to the network, but may never end up doing so,
due to its parochial view.

TCP has been designed to work in a broad range of networks. Each TCP
variant works well for some kinds of the network and its traffic and
the same TCP performs poorly for other conditions. The interesting
part is that, we do not know exactly what objective does TCP
congestion control try to optimize \cite{Winstein}. This inflexibility
in adapting to new scenarios limits its use. 


The emergence of Software Defined Networking (SDN) gives network
protocol designers the power of centralized view and centralized
control that can be exploited for many applications
\cite{mckeown,SDN}.  SDN provides a centralized view of the network
with access to the statistics and other information of the routers and
link states. The central controller can aggregate these information
and actively participate in the congestion control decisions of the
end nodes.  The scope of this paper is to study to what extent we can
exploit the central view and the centralized control features to
improve congestion control in networks.

An SDN-enabled scheme for handling congestion control is presented in
the paper.  With a global view, the controller knows exactly what each
of the link states are and would never misinterpret a packet error as
congestion (as was done by many TCP variants). The information at the
controller can supplement the indicators like packet loss and delay,
that were used by the end nodes earlier. The controller can provide a
more realistic view of the network to the end nodes.  With a more
detailed knowledge about the network and the traffic flowing at any
point in time, we can take better, faster congestion control
decisions. In the proposed mechanism, the controller instructs the
switches (via the OpenFlow API) to set the relevant ECN bits on
packets going through a switch. This information is then used by the
TCP end-nodes for changing the TCP congestion window. The scheme has
been implemented in the Mininet emulator \cite{mininet} and studied
for three different network scenarios. The results show that the
proposed approach achieves improvements over TCP CUBIC and TCP/RED
based distributed solutions.



\section{Background and Related Work}

This section presents the relevant background material and related work.

\subsection {Router-based Congestion Control}

The DECbit mechanism \cite{Ramakrishnan} is one of the first works
that used an explicit congestion control protocol to signal the end
nodes about congestion at a router. An router, when it is likely to
experience congestion, reacts by marking a bit at the packet header
for some of the packets. This bit was then sent by the receiver to the
sender for taking appropriate action.

The Random Early Detection (RED) mechanism \cite{FloydRED} attempts to
maintain an average queue length at the routers.  Using threshold
values for time-averaged queue lengths, a router drops packets with a
probability that increases with the size of the queue. This packet
drop results in TCP sources reducing their congestion window and hence
transmission rate.

In the Explicit Congestion Notification (ECN) mechanism \cite{Floyd},
packets will be marked with Congestion Encountered (CE) bits instead
of being dropped. The end-node TCP protocol is modified so that the
receiver echoes the CE bits to the sender. When the sender receives
such TCP segments, it reacts by reducing the congestion window.  TCP's
performance can be increased significantly using RED/ECN by setting
the parameters with caution. The DCTCP protocol \cite{alizadeh} is a
variant of TCP optimized for data center networks.  DCTCP uses
Explicit Congestion Notification (ECN) at the switches and runs its
variant of TCP at the end nodes. In \cite{wu}, another ECN-based
scheme that does not require end-host TCP modification is presented
for data center networks.

In all these approaches, the decision to drop or mark packets during
congestion is taken in a distributed manner by each router.  In an
SDN-enabled network, it is possible to take advantage of the
(logically) centralized control plane to set the bits based on the
global view.  This paper presents an attempt to explore the advantages
of pushing this decision to the SDN controller rather than doing it at
the routers.

\begin{figure*}[!t]
\centering
\includegraphics[width=4.5in]{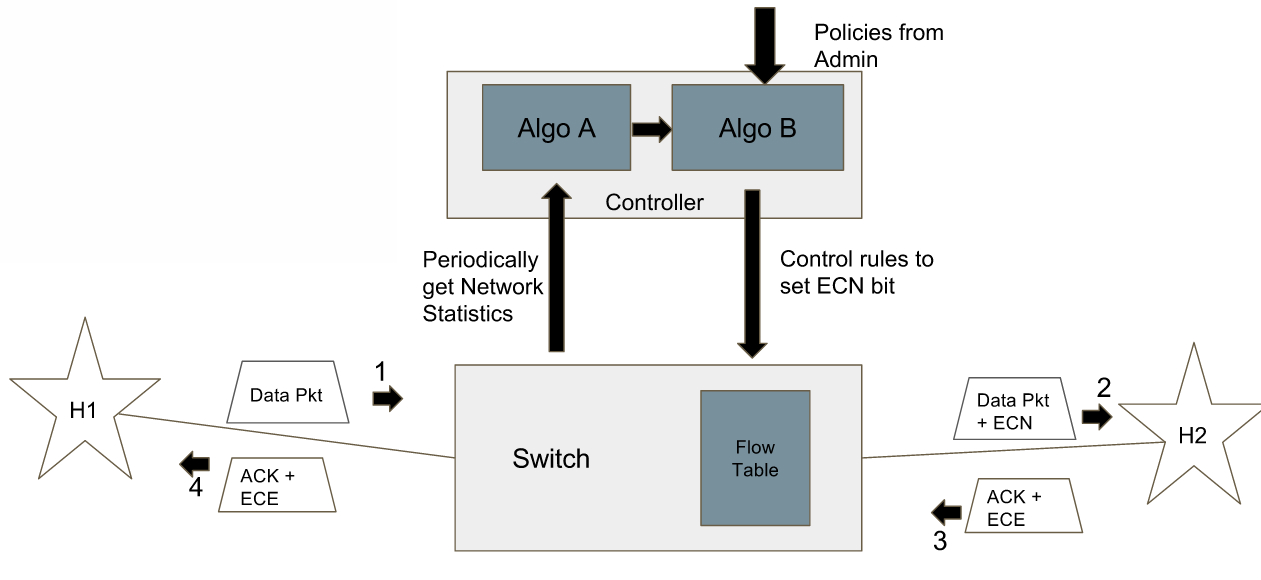}
\caption{Proposed SDN-based Congestion Control Framework.}
\label{fig_Arch}
\end{figure*}

\subsection{OpenTCP}

OpenTCP \cite{Ghobadi} is a congestion control framework proposed for
SDN based networks.  The OpenTCP software at the controller collects
information about the underlying network such as link utilization
values. OpenTCP aims at reducing the Flow Completion Times (FCTs) by
updating initial congestion window and retransmission timeout interval
for TCP flows. OpenTCP sends Congestion Update Epistles (CUE) packets
to the end nodes consisting of information about the suggested changes
to their TCP state. OpenTCP's kernel module running in the TCP stack
of the end node updates their TCP variant's state using the
information in the CUE packets. Thus, OpenTCP requires the end node
TCP protocol to be changed.

The approach proposed in this paper is SDN-based and using the ECN
mechanism, but does not require any changes to the end nodes' TCP
protocol. This is advantageous since changing end nodes may not be
possible if all the nodes are not under the same administrator,
especially in multi-tenant networks and data centers providing cloud
services.

\section{Proposed SDN-based Framework}

The proposed SDN-enabled framework is shown in
Fig.~\ref{fig_Arch}. The network consists of several switches/routers
that implement the \textit{data plane} with the \textit{control plane}
implemented by a logically centralized SDN controller.  The end-nodes'
TCP protocol implementation is not modified, but is expected to be
ECN-enabled.  An SDN controller application periodically collects
information about the underlying network and the traffic
characteristics. Based on these information, it detects congestion
according to an algorithm (generically called {\bf Algorithm-A}).

With the global view of the network available, another algorithm
(called {\bf Algorithm-B}) selects the end nodes that need to react to
any detected congestion. This algorithm also considers the policies
and the priorities to flows set by the administrator to decide which
end nodes to penalize. After taking these decisions, the application
sends new flow rules to the switches (which timeout after a transient
amount of time), instructing them to set the ECN bits for particular
flows. The application conveys information regarding congestion to the
TCP end hosts by setting ECN bits. The normal TCP end nodes, react to
packets marked with ECN bits by reducing their congestion
window. Different implementations of \textit{Algorithm-A} and
\textit{Algorithm-B} can be used for different types of networks and
workloads.

\subsection{Congestion Detection} The congestion detection component
(\textit{Algorithm-A}) periodically collects information about the
network and its traffic and tries to recognize/predict congestion. SDN
switches have the capability of collecting statistics. The application
can probe the switch for these statistics and get a complete picture
about the network and its present traffic conditions. Queue statistics
and link utilization statistics will be used by \textit{Algorithm-A}
for detecting congestion. A simple metric to identify potential
congestion is if the link utilization or average queue length is
greater than a threshold. More sophisticated algorithms could consider
global network state and predict congestion based on learning
algorithms or other complex heuristics. In the evaluation section, we
will present an example for \textit{Algorithm-A}.

\subsection{Handling Congestion} 
Once \textit{Algorithm-A} predicts congestion in a link/switch,
\textit{Algorithm-B} decides on how to handle the congestion.  The
administrator can also set policies for congestion control. For
instance, the policy can specify that certain type of high-priority
flows should not be affected.  It will compute different ways to avoid
congestion, based on the global view and administrator policies. It
will identify the set of end nodes whose congestion window can be
reduced to avoid congestion while maximizing a global objective.

In traditional TCP, all the end nodes react to congestion by reducing
their congestion window. In data center networks and other SDN
applications, it is not necessary that all the flows should react to
congestion. For instance, we would like the interactive traffic or
short-term (i.e. \textit{mice}) flows to be not modified but control
the congestion windows of bulk (i.e. \textit{elephant}) traffic. More
over, when all the nodes react to congestion, they all back off at the
same time (leading to under utilization) and then move forward
simultaneous giving rise to a saw-tooth like bursty traffic at the
switches. 

For example, the following scheme can be used as \textit{Algorithm-B}
in data center networks for prioritizing interactive flows.  Whenever
congestion is detected in a link, the framework will obtain all
statistics about the flows traversing that link. The algorithm can
select the top $T\%$ of the the flows based on their bandwidth
utilization and reduce their congestion windows. Generally, only about
10\% of the flows are elephant flows in the network, but they utilize
about 90\% of the bandwidth. Penalizing the high utilization flows can
ensure that interactive traffic is less affected by congestion.

\subsection{Congestion Notification}

The next step is to convey the congestion handling information to the
end nodes so that they can react appropriately. The
\textit{Algorithm-B}, after computing the set nodes that have to
reduce their congestion window, will send explicit flow rules to the
switches with an additional action of marking the ECN bits apart from
forwarding the packets to the right port.  These new high priority
rules sent by \textit{Algorithm-B} are called the Congestion Control
Flow Rules (CC flow rules). When new packets from these flows arrive,
they get matched with these top priority CC flow rules and their ECN
bits get set at the switch before forwarding. 

The ECN bit processing is based on the end-nodes' TCP
implementation. As mentioned earlier, we do not require any changes in
the TCP implementation.  A receiver, upon receiving packets with
marked ECN bits, echoes the information to the sender by marking the
ECE bit in their ACK packet. The sender acts upon these marked ACKs
and reduce their congestion window according to their native TCP
variant.  

The new ECN marking flows added are given higher priority in the
OpenFlow tables at the switches, compared to the regular flows. This
is to make sure they are not skipped. Also, these new CC flow entries
are set with a very small rule timeout interval so that adequate
number of packets are marked. If a new CC flow rule is added for a
very long time, it might send signals of large congestion to the end
node that would affect the throughput of that flow adversely. At the
same time, the timeout interval should not be too low such that not
enough packets from that flow actually get marked. Choosing this
timeout value is a critical parameter. Once the timeout interval is
over, these extra high priority flows get evicted and the flows start
matching with the normal low priority flows rules that were already
present in the switches.

Another critical parameter is the periodicity of the network probes.
As mentioned earlier, \textit{Algorithm-A} probes the network for
network and traffic state. There should be enough time for the system
to settle down after the congestion window changes are implemented by
the source TCP and the network traffic stabilizes. The ECN packets
have to reach the receiver and the receiver has to reply back with
bits marked in the ACK packet. It clearly takes about 2 RTT to get
communicated to the end nodes and it might take a couple of more RTTs
for the network to become stable. Very large probing interval cannot
detect sudden congestion and can be less efficient. Thus, the probing
interval is a sensitive parameter.

In summary, the salient features and advantages of this framework
include:

\begin{enumerate}
\item {\bf Global View:} This enables the congestion control scheme to
  obtain a more complete picture about the network state and hence
  make better congestion predictions that can achieve more globalized
  objectives.  In conventional end-to-end systems, all the end nodes
  try to optimize their own local objective function and could end up
  in a Nash equilibrium solution.

\item {\bf Prioritization in Congestion control:} Flow priority can be
  considered in making congestion control decisions and in setting of
  flow table rules.

\item {\bf Fairness:} Fairness can be ensured even across TCP and
  UDP. Typically, UDP flows can hog the bandwidth placing TCP flows at
  disadvantage. With the proposed approach, we can monitor the
  bandwidth used by UDP and use special flow rules to restrict UDP's
  uneven share of the bandwidth.

\item {\bf No change to the end-nodes:} The proposed approach does not
  require changes to the end-nodes' protocol stack.

\item {\bf No change to switches:} This approach uses normal flow
  rules and the functionalities available with existing SDN switches
  to achieve congestion control.

\item {\bf Easily pluggable CC algorithm}: The Algorithms $A$ and $B$
  can be changed according to different network and traffic needs. A
  data center network may need a different kind of algorithm than an
  enterprise network. Congestion control algorithm at the controller
  becomes a easily changeable mechanism.

\end{enumerate}

\section{Performance Evaluation}

This section presents the performance evaluation of the proposed
congestion control framework and compared to existing schemes.

\subsection{Implementation Details}

The proposed framework has been implemented in an SDN emulator package
called Mininet (version 2.2.1) \cite{mininet}. The Floodlight SDN
controller \cite{floodlight}, Open vSwitch (version 2.3), OpenFlow 1.3
that supports setting of ECN bits through flow rules have been used.
The congestion control framework has been implemented as an
application in Floodlight. In all our experiments, the switch contains
a single flow table. All the virtual end hosts created by Mininet in
our experiments run TCP Cubic implementation that comes with Ubuntu
14.04 kernel. The ECN mechanism has been turned on at all the end
nodes. Since Mininet is not able to handle high bandwidths accurately,
100Mbps links have been used everywhere unless stated otherwise.

The analysis is done for data center networks with the objective of
achieving lower flow completion times (FCT) for interactive traffic
when it co-exits with bulk traffic.  As a proof of concept for our
framework, we present sample congestion control algorithms,
(\textit{Algorithm-A} and \textit{Algorithm-B}), and evaluate their
performance.

The proposed \textit{Algorithm-A} probes the switches every 2 seconds
and collects port statistics information. From these, we compute the
average link utilization of every link in that interval. If the link
utilization is greater than 75\% , we consider that congestion is
likely to occur in that link and inform \textit{Algorithm-B} about
it. The handling algorithm, \textit{Algorithm-B}, is defined as
follows. Congestion control is imposed on the top $T$\% of flows
(based on bandwidth utilization) going through the congested link once
the link utilization is greater than 75\%. We linearly increase $T$ as
link utilization increases and when link utilization reaches 100\%, we
set $T = 50\%$ penalizing top 50\% of the flows when utilization hits
the maximum. Since we are not penalizing all the flows at the same
time, this can prevent saw tooth like behavior of
utilization. Additionally, we start penalizing more and more as we get
closer to 100\% utilization making sure we have enough bandwidth for
new incoming short flows.


The proposed approach is compared with Linux's TCP Cubic scheme, and
approaches that use in-network elements in congestion control, namely
RED \cite{FloydRED} and ECN \cite{Floyd}. The implementation of RED
and ECN is available in Mininet. 

\subsection{Throughput with long flows} The objective of this
experiment is to show that the proposed algorithm does not compromise
on bulk flows and achieves good throughput on long-lived traffic.

The topology studied, denoted \emph{Topology~1}, consists of one switch
connected to four end nodes with 100 Mbps links each. Three nodes are
senders ($S_1,S_2,S_3$) and one is a receiver ($R_1$).  Three of the
senders run \emph{iperf} for two minutes to generate traffic to the
receiver. The results are shown in Table~\ref{throughput_bulk}. As
seen, the proposed approach gives 8\% better total throughput than
Cubic and comparable performance with the distributed ECN and RED
schemes. The proposed method achieve better fairness than TCP Cubic
and is on par with ECN and RED.

\begin{table}[!t]
\renewcommand{\arraystretch}{1.3}
\caption{Throughput of bulk flows for Topology 1 (Mbps)}
\label{throughput_bulk}
\centering
\begin{tabular}{|c|c|c|c|c|}
\hline
Flow & TCP Cubic & ECN & RED  & Proposed \\
\hline
S1 &32.5 &29.6 &30.9 &32.7 \\
\hline
S2 &25.8 &29.4&28.1&29.7 \\
\hline
S3 &26.0 &30.2&32.2&28.7 \\
\hline
Total & 84.3&89.2&91.2&91.1 \\
\hline
\end{tabular}
\end{table}

The next topology, called \emph{Topology~2}, is a dumbbell topology
with 2 switches and 3 nodes connected to each switch. Nodes connected
to one of these switches are all senders, called $S_1, S_2, S_3$. The
nodes connected to the other switch are receivers $R_1,R_2,R_3$.
Three pairs ($S_1,R_1$),($S_2,R_2$),($S_3,R_3$) are selected and
long-lived flows are established between them for 2 minutes. All the
links have 100 Mbps capacity. The throughput results are shown in
Table~\ref{throughput_interactive}. As seen, the proposed approach
achieves 14\% better total throughput than TCP Cubic and is similar to
that achieved by RED and ECN. The individual through-puts are also
closer to the fair-share values.

\begin{table}[!t]
\renewcommand{\arraystretch}{1.3}
\caption{Throughput of bulk flows for Topology 2 (Mbps)}
\label{throughput_interactive}
\centering
\begin{tabular}{|c|c|c|c|c|}
\hline
Flow & TCP Cubic & ECN & RED &  Proposed \\
\hline
S1 &27.5&31.3& 32.9&31.3\\
\hline
S2 &33.3&29.9& 30.9& 31.9\\
\hline
S3 &23.3&32.9& 32.1&32.5\\
\hline
Total (Mbps)&84.1&94.1& 95.9&95.7 \\
\hline
\end{tabular}
\end{table}

\subsection{Coexistence of interactive and bulk traffic} In this
set of experiments, we show that the proposed algorithm significantly
improves the flow completion time of interactive flows in presence of
bulk traffic. In the first experiment, we consider the single
switch-four node \emph{Topology~1}.  Two long-lived flow between from
$S_1$ and $S_2$ to the receiver $R_1$ are established. A 2~MB
interactive flow is sent from $S_3$ to $R_1$. This experiment is
repeated 30 times for each of the congestion control algorithm and the
interactive flow's mean flow completion time is computed, as shown
below.

\begin{center}
\begin{tabular}{|c|c|c|c|c|c|}
\hline
TCP Cubic & ECN & RED &  Proposed \\
\hline
10.64 & 0.36 & 0.41 &  0.34 \\
\hline
\end{tabular}
\end{center}

As seen, the proposed scheme obtains 30x improvement to Cubic in flow
completion time and about 1.2x improvement to RED. ECN was close (but
worse) to our scheme in this topology. We also found that our
algorithm shows the least variance in flow completion time compared to
the other schemes, based on the 30 iterations.

The \emph{Topology~2}, discussed in the previous section, was next
considered and flows established as above. The flow computation time
results based on 30 iterations is shown below.

\begin{center}
\begin{tabular}{|c|c|c|c|c|c|}
\hline
TCP Cubic & ECN & RED &  Proposed \\
\hline
 10.41 &0.46 & 0.37 &  0.37\\
\hline
\end{tabular}
\end{center}

As seen, the proposed approach performs 28x better than TCP Cubic,
1.25x better than ECN, but on par with RED. It was observed that the
proposed schemes produce had the least variance among the four.

To summarize, the proposed scheme outperforms TCP Cubic by a large
factor. It works the best for both the scenarios, while ECN and RED
failed to perform better than the proposed scheme in at least one of
the cases.

\begin{figure}[!t]
\centering
\includegraphics[width=3.5in]{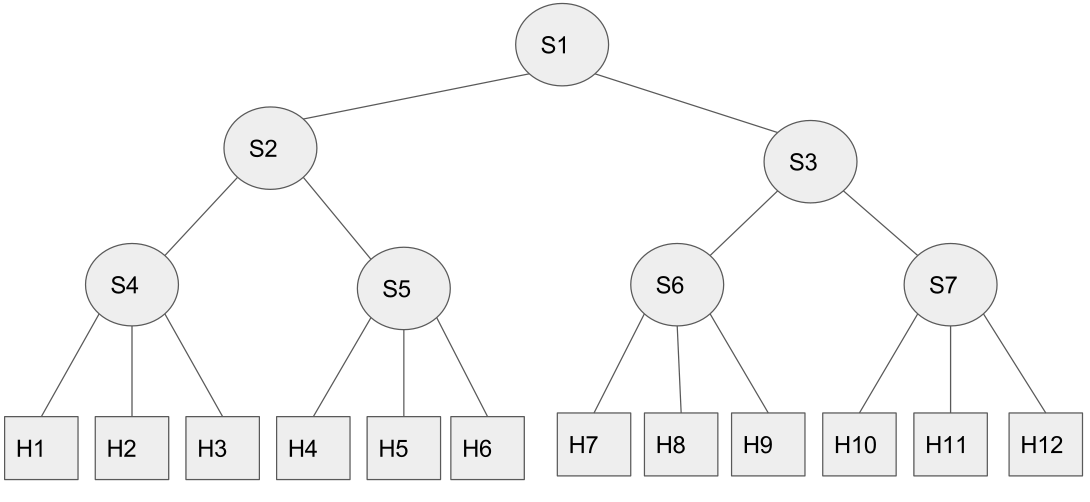}
\caption{Multi-hop Topology}
\label{fig_top}
\end{figure}

\subsection{Multi-hop scenario} 

We evaluate the proposed algorithm in multi-hop networks with multiple
bottleneck links. The 7-switch topology is shown in
Figure~\ref{fig_top}.  In the first experiment, we establish long
lived flows from $H_1,H_2,H_4,H_5$ to $H_7,H_8,H_{10},H_{11}$
respectively for 2 minutes. The total throughput (in Mbps) is
presented below. As seen, the proposed scheme achieves throughput
better than TCP Cubic and comparable to ECN, but less than that of
RED. 

\begin{center}
\begin{tabular}{|c|c|c|c|c|c|}
\hline
 & TCP Cubic & ECN & RED &  Proposed \\
\hline
Throughput & 76.7 & 83.9  & 85.4 & 83.8 \\
\hline
\end{tabular}
\end{center}
 
To check flow completion time for mice flows in this setting, We
created long-lived flows from $H_1,H_2,H_4,H_5$ to
$H_7,H_8,H_{10},H_{11}$ respectively. We simultaneously sent 2MB data
from $H_3$ to $H_7$ and from $H_6$ to $H_8$ and measured the flow
completion times. This experiment was repeated 30 times. The average
flow completion time (in seconds) achieved by each of the algorithms
are shown below. 


\begin{center}
\begin{tabular}{|c|c|c|c|c|c|}
\hline
TCP Cubic & ECN & RED & Proposed\\
\hline
19.46 & 1.01 & 1.1 & 0.64 \\
\hline
\end{tabular}
\end{center}

In multi-hop conditions, our scheme clearly outperforms ECN by a
factor of 1.5x and RED by 1.7x without compromising much on throughput
of bulk flows. Our scheme also outperforms TCP Cubic by 30x.







\section{Conclusions}

This paper has presented a simple and easy to deploy framework for
congestion control in Software Defined Networks. The framework is
extensible in terms of specific congestion control and detection
algorithms. It requires no changes to the end nodes or the SDN-enabled
OpenFlow switches. The proposed framework has been implemented in the
Mininet emulator, with heuristics for data center networks. The
proposed approach shows 30x improvement to TCP Cubic, 1.7x to RED and
1.5x to ECN on multi-hop topologies for flow completion times of
interactive traffic without compromising throughput of bulk flows.

\ifCLASSOPTIONcaptionsoff
  \newpage
\fi

%

\begin{IEEEbiography}{Michael Shell}
Biography text here.
\end{IEEEbiography}

\begin{IEEEbiographynophoto}{John Doe}
Biography text here.
\end{IEEEbiographynophoto}


\begin{IEEEbiographynophoto}{Jane Doe}
Biography text here.
\end{IEEEbiographynophoto}




\end{document}